# Shared Services Center for E-Government Policy

Flavio Corradini*, Lucio Forastieri**, Alberto Polzonetti*,
Oliviero Riganelli*, Andrea Sergiacomi**

*Univ. of Camerino – Dip. MATINF – Via Madonna delle Carceri, 9
62032 Camerino (MC) Italy
+39 0737 402564 - +39 0737 402561
{flavio.corradini, alberto.polzonetti, oliviero.riganelli}@unicam.it

** Regione Marche Servizio Informatica Via Gentile da Fabriano, 9
60125 Ancona (AN) Italy
+39 071 8063500 - +39 071 8063071
{lucio.forastieri, andrea.sergiacomi}@regione.marche.it

**Abstract:** It is a general opinion that applicative cooperation represents a useful vehicle for the development of e-government. At the architectural level, solutions for applicative cooperation are quite stable, but organizational and methodological problems prevent the expected and needed development of cooperation among different administrations. Moreover, the introduction of the "digital government" requires a considerable involvement of resources that can be unsustainable for small public administrations. This work shows how the above mentioned problems can be (partially) solved with the introduction of a Shared Services Center (SSC).

## 1 Introduction

One of the most challenges of the e-government paradigm, also according to the available technology, is to integrate information systems in such a way that different administrations can suitably interact each other by sharing activities, processes, services etc. This integration reduces citizens' time dedicated to bureaucracy, increases quality of services, allows to share knowledge, information and best practices [BCSV03, DEEEH03,FRI02].

Since 1993, several Italian initiatives have been considered by both the central government and regional ones to support the integration of the citizens and the state. An effective service to citizens relies on a cooperation at the applicative level among the various administrations; in other words we need a sort of distributed informative system where independent public administrations (PA) are involved and share services each other [MECBAT01a].

The technologies that support a Cooperative Information System (CIS) are mature enough to be fruitfully used in a large class of applications. The same, however, cannot be said for the underlying developing methodologies [MECBAT00]. The crucial point is to develop a general architecture to integrate systems that have been designed to support vertical applications. Hence, it is needed a re-engineering process of procedures and services.

The reorganization of informative systems in an organization needs a parallel reorganization of both administrative and economic processes with constraints concerning "legacy" architectures, organization structure and actual laws. In general, hence, we need an architecture that coordinates the exchange of information among the various information systems by leaving full autonomy to each organization together with a cyclic and iterative methodological process that takes into account the constraints imposed by the different organizations, the different cultures and the different internal management [DAWPRE03, KRAZEM01, MECBAT01b]. To this aim have been defined guide lines for interoperability services and applicative cooperation [AIP00, DUR03, DUR04, MIT03a, PCM00, VIR01].

Fig. 1 shows a three layered cooperative architecture that offer connectivity, base services and cooperation services. In these last years several pilot projects have experimented potential solutions (middleware-based, web-based and traditional) to this problem. The digital government, so that other areas of ICT, is dominated by the Web. The web services qualify the public administrations to provide services by means of the definition of new services (even from other e-government sources) through the standardization of the description, the research and the invocation of cooperative procedures [ELMMCI01].

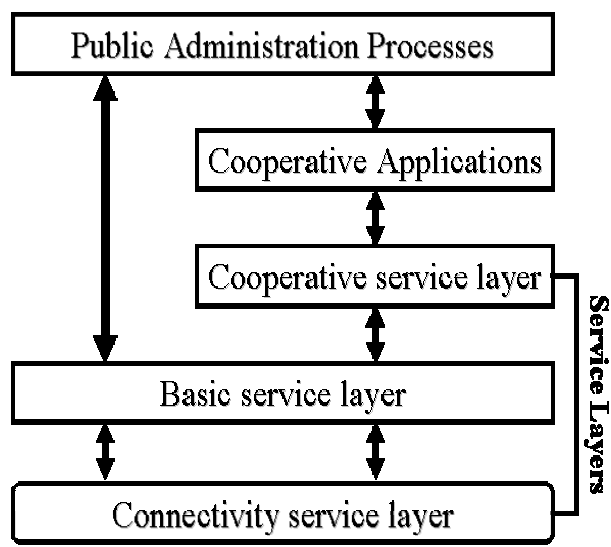

**Fig. 1.** Service layers

The e-government digital services, indeed, classify as one of the most intriguing areas in the Web Services setting because of the innumerable factors (human resources, geography, ontology organization, security, quality of data etc.) that make this domain definitely more complex than others [AFNT04, ELMMCI01]. For these reasons, this work concentrates on the concept of "shared service center" as useful tool that let various PA to share those functionalities that are in common and that is of difficult realization and management. The more complex functionalities are phisically located in some organizational structure able to:

1. take part to the digital government to those organizations that are not able to interoperate,

2. allow the public access to services through a unique interface, possibly without considering whether these services are provided by a single or several organizations. One-stop government reflects a new concept of PA (see Fig. 2) based on the way the PA's distribute their own services [WIMKRE01].

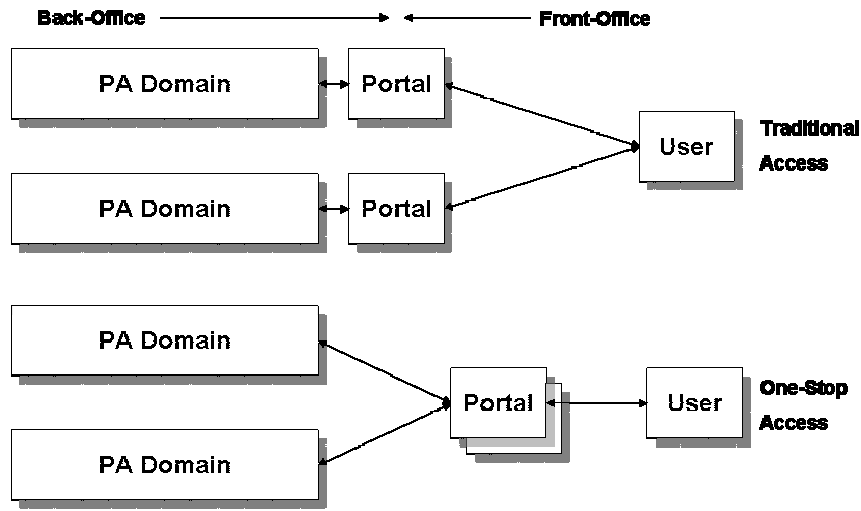

**Fig. 2.** On-line one-stop government

## 2   Migrating ICT outsourcing to Shared Services Center

Public administrations are facing a very complex renewing process within their organizational structures trying to improve internal working modalities and the quality of the given services. Two main particular phenomenon contribute to favor such a modernizing. On one hand the introduction of the information and communication technologies (ICT) radically changes the way employees work together with the given available services; on the other hand, the reorganization of some functions through outsourcing processes[BIL01]. Evolution of available technology, economic pressure and needs to canalize suitable resources on the "core competencies" (meant as set of strategic activities for the PA)  make the management and the development of ICT products by administrations can be very difficult. In order to face these problems some administrations resort to outsourcing by committing parts of the services to external partners. This allows to optimize the use of human and technological available resources and to speed up the distribution of  e-services and the information for citizen and firms.

While in the past were mainly considerations about costs in favor of ICT outsourcing, in these last years the process view [LHKP03] is becoming more and more significant. The external partners are not only "service providers" but directly integrated into the "value chain". This integrated framework allows the PA both to evaluate the social impact and to detect strategies connected with the operation. Independently from the choice make-or-buy, potentially convenient, the main aim is to find a proper  partnership between the administration and the external offer that, in turn, means to optimize concurrency management, costs and the quality in favor of the involved administration [PAU03].

Outsourcing for administrations is more than cost-effective. Function delegation has allowed, and actually allows, to public organisms to have a unique interface for the solution of a number of complex problems and to have the possibility to take part in a direct and immediate way to exceptions that can happen. This allows to continuously improve the "services standard" and  to reach a low level of bureaucracy connected with the given functions. A Shared Services can be seen as a particular king of outsourcing between several clients and a single provider. In particular, SSC is defined as a concentration of resources, detected as activities that are typically distributed crosswise to the organization with the main aim to serve different users at low costs and with a high level of service. This has a direct consequence to satisfy external users and increase the corporative image [SHDL99]. Hence,  we can further distinguish between "ICT outsourcing" and Shared Services: the former represents a legally service provided by an independent third part, while the latter is a proper function within a "corporate group" [ULB03]. The idea of SSC is then to gain from the investments in the domain of e-government by sharing common elements present in the single administrations.

| | |
|---|---|
| **Strategic and organizational benefits** | The local units focus on their core administrative process |
| | Clear control and eliminate local and complicated control of IT function |
| | Gain access to high quality IT services and skills |
| | Concentration of innovation and exploit new Technology |
| | Share risks |
| | One-stop shop |
| | Standardize functionality and processes among administration |
| | Disseminate and impose successful practices |
| | Reduction of complexity |
| **Political benefits** | Enhance credibility |
| | Solve internal conflicts |
| | Increase controllability |
| **Technical benefits** | Concentration and access to technical and project management expertise |
| | Poor performance of local ICT staff |
| | Higher services levels |
| | Consolidate experiences |
| | Escape from legacy systems |
| | Standardization of platforms and application Vendors |
| | Better information security and authorization by centralizing |
| **Economical benefits** | Lower control and maintenance costs |
| | Accountability of control |

**Tab. 1** . Benefits on using a Share Services environment

In a shared services environment, the PA externalizes the activities that support their core competencies by organizing them in separate functional modules. As a consequence, the single PAs can concentrate their attention to the development of the most strategic activities for the administrative process. By developing a methodology based on SSC, moreover, two communities with different competencies can be isolated: the specialists for the design and realization of Share Services, that provide their experience in the technology and projects management, and the local experts that contribute to provide the needed knowledge of the administrative process and user requests [JANWAG04].

## 3   Analysis of the Share Services Infrastructure

SSC aims at supporting the technical aspects of management and distribution through telematic means of the administrative services based on a number and heterogeneous data and knowledge, data bases and services, given by the different administrations. The existence of an infrastructure supports the realization and put forward administrative services for citizens and firms, in particular for those entities that cannot manage, in full autonomy, the distrubution via telematic means by a lack of both economic resources and planning skills. The infrastructure includes a framework (see

Fig. 3) that provides a series of solutions for complex problems and a set of shared and standardized services that can be used according to the usual mechanisms of the applicative cooperation [PCM00].

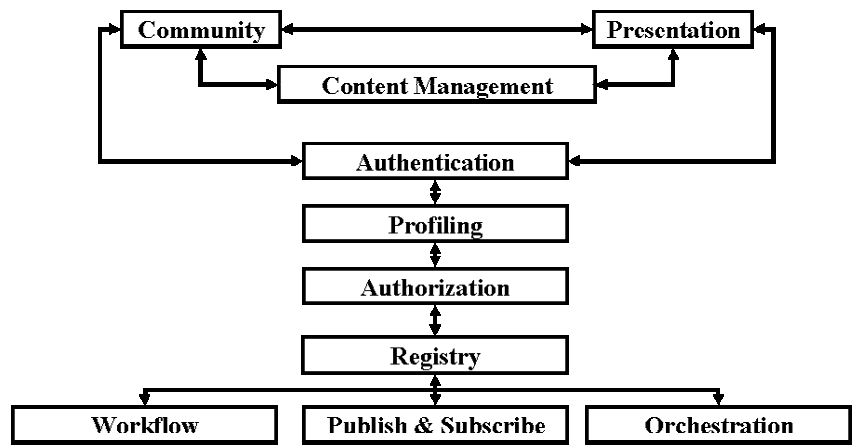

**Fig. 3.** General Framework

In particular, the framework provides services for applicative cooperation, for security and the Single Sign On, for the automation of the procedural and documental flows, for content management. This last aspect allows a significant rationalization of the content management platforms usage, the relative access modalities and information updating, the management and maintainability of informative services. The shared services are provided by means of a modular architecture that is based on "building blocks" reuse. The services are exposed via web services and through a modality of interaction with non invasive back-office systems. The possibility to develop services among the different PA, in a G2G logic, allows to increase efficiency in administrative procedures and to reach the objective to present the different PA to citizens and firms in a more and more global perspective (G2C,G2B). The reference model is based on the distribution of services in local and regional portals that constitute the telematic front-office for the users. While the presentation of services can be suitably differentiated and personalized in function of the access typology on the user side, the given services, instead, remain a shared patrimony controlled in the access modality and checked via the services catalog.

The framework is logically located between the front-office and back-office as part of the infrastructure of the applicative cooperation and located in the middle in the distribution of services based on direct interaction (synchronous) between delegated and applicative ports in the front-end and back-end systems. This mediation plays a role when the interaction relies on some asynchronous mechanism of applicative cooperation as, for instance, those based on Publish&Subscribe and process orchestration. We can now give a list of services needed for the above mentioned functions:

**Internal Services** for the management of the technological infrastructure. These are centralized and demanded to specific operative points that also monitor the systems and the applications, control the security conditions and, in general, the operative and administrative applications needed for the correct operativity of the informative system.

**Services for the applicative cooperation** based on synchronous and asynchronous collaboration profiles for the exchange of the informative flows between delegated and applicative ports coherently to the model of the e-Government envelop:

- *Process Orchestration*, via a central engine able to manage instances of registered models in a suitable repository; significant knowledge base for the diffusion of process logics and for the reduction of analysis time,
- *Event notification systems*, based on Publish & Subscribe that guarantee a strong decoupling between the parts in terms of distribution, timing and flows of events,
- *Workflow services*, for automation of documental flows and the management of complex processes that require, for the completion of one or more steps, also the interference of "real" users and not only applicatives
- *Security Infrastructure*, in order to guarantee attributes as integrity, confidentiality and non repudiation of exchanged messages between delegated and applicative ports.

**Registry Services**, for registration and consulting. The services catalog must be able to host in their archives the information that concern the given applicative services and the tools that let them available to the community.

**Logging services**, for the traceability of services requests, with accounting of the orchestration state or of the workflow processes and in general for the management of generic applicative events, including errors and exceptions.

**Authentication and authorization services,** for the centralized management of the accesses to the reserved areas of the portals exposed in the framework

- *Single Sign On (SSO),* for the transfer of credentials between authenticated users between access portals; the authentication on the framework is possible via weak and strong registration, or via the use of a card for the digital signature or via services cards.
- *Profiling systems*, for the coordinated management of information on the users with credentials, logically divided in a static subsystem (extended registry) and a dynamic one, containing a series of attributes able to indicate preferences of the user when accessing the services rather than informative areas on portals.

**Content Management services**, able to manage in a centralized manner the contents on access portals and their automatized distribution on external servers:

- *Front-End Services*, for the integrated management of both static sites and personalized templates, populated by contents and services filtered with respect to the preferences in the user profile,
- *Community services*, for the simplified management of the News and Forum areas of informative portals.

## 4   Case Study: "Regione Marche" Shared Services Center (RMSSC)

In Italy several initiatives have been undertaken to explicitly support small municipal districts in order to guarantee that citizens can participate in the e-government innovation processes. There are 5.836 small municipal districts of about 5000 citizens each. Three kinds of difficulties characterize such areas:

- The impossibility to get proper funds for the realization of innovative processes,
- The lack of adequate skills to support innovation,
- The lack of a proper technological infrastructure.

In spite of such difficulties, that risk to get out the smaller administratives from the actuation of e-government and, at the same, it makes even worse the negative phenomenon of the territory marginality, some excellence realities just located in small municipal districts, show a live participation to the e-government processes. This guarantees equal opportunities for all citizens, independently from where they reside, but also the inclusion of small public administrations in the innumerable opportunities offered by ICT [MIT03b].

In the actuation of the national e-government plan, the "Ente Regione Marche" has a coordination role both at the planning level, to detect a proper strategies within the territory, and at the management level because it has the authority to manage the cooperation among underlying entities (such as Provincie and Comuni). In the context of the experimentation, the Ente Regione has a significant role and is actually involved in a remarkable technological upgrade that has supported the design and implementation of a Shared Services environment (see Fig.4) [RGL03]. The experimentation involves around the 80% of the municipal districts of the Marche region and relies on a collaboration between Regione Marche, Univerità degli Studi di Camerino and the involved districts.

Our experimentation uses an infrastructural platform of digital community based on an organizational system of standards by means of a Shared Services Center methodology. Two main activities are considered: the first related to the "Thematic Map" of the citizens and firms services and the second related to the "one-stop government". The delivery of on-line services to citizens and firms is mainly through institutional web sites or through aggregated portal of services. There is a variety of public portals that do not guarantee any quality issue, reliability and interaction homogeneity with the public administration, often without any check against the users perspective. These problems have the immediate effect of discouraging and preventing the users with respect to the new technologies and the new services.

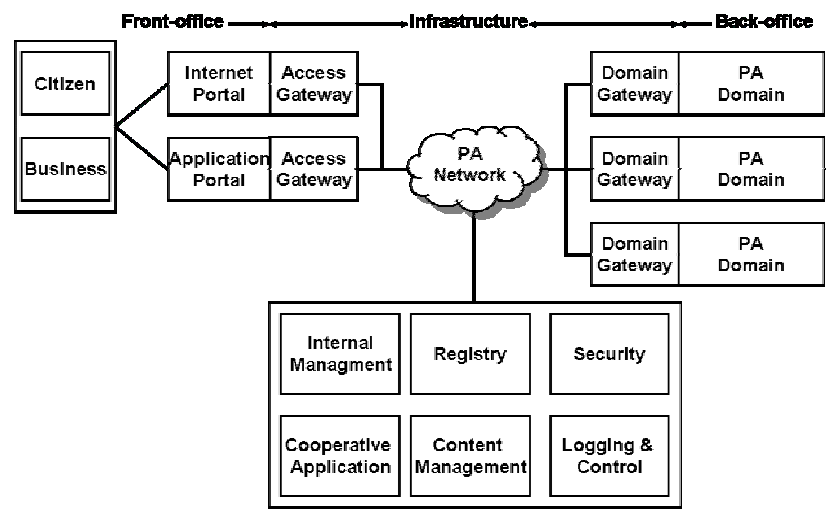

**Fig. 4**. Shared Services infrastructure

We aim at optimizing the investments done by the single PA for the realization of their own Front-Office services. For such a reason it is needed to define a common on-line interaction model among PAs that give value to the internet visibility, the specificity of the services provider and make ensure at the same time interoperability and service provider identification. By using the "registry" functions of the SSC we provide the activities that allow the authorized entities to put forward their own services according to a taxonomy of "life events" and usage target. From the perspective of the user, this allows to rely on a unique tool in order to know the available services and use them from different profiles. We have already realized a unique identification system or, better, an identification system that allows to use any service from any provider by exploiting the current technology on authentication, authorization and single sign-on while the organization on the territory of authentication flows are under design and development. The execution of several administrative procedures that require the interaction among several PAs has needed the development of a second activity. Currently, it is often required that citizen or firm directly undertake the integration of the back-office services of the different involved administrations; the activities consist of the automation of processes that give a unitary and global view of the public administration from the final user perspective. This concept is better known as one-stop shop or one-stop government. The realization of such an activity is difficult because of the lack of shared back-office services that is prevented by the already mentioned problems regarding culture, geographic location and number of different italian public administrations. On the other hand, the lack of shared back-office services has allowed the starting of procedures and services re-engineering process needed to allow an horizontal applicative cooperation among PAs.

## 5  Concluding remarks and future works

This work shows as the evolution of "ICT outsourcing" in "Shared Services" can significantly support the development of e-government processes. The applicative cooperation represents one of the most involved problems of the digital government. The aim of our proposal is twofold. On one hand it allows to overcome the deficiencies concerned with the lack of a proper know-how within small public administrations. On the other hand, the Ente Regione as Shared Services Center, enforces the territorial competencies of the Region itself by allowing a significant cost-effective use of technologies. We have realized a highly technological supporting framework that has already been validated over some given functionality. We plan to validate the framework over more complex functionalities such as workflow, publishing and subscribe and orchestration in a territorial context that is very much split at the administrative level. Moreover, we have also planned to verify the features of the realization of the one-stop government to make it uniform with respect to the standard in the literature [KRE02, TAM01, TRAWIM01, WIMKRE01, WIMTAM02] and to guarantee a wide applicability.